# Thin film synthesis of semiconductors in the Mg-Sb-N materials system

Karen N. Heinselman *[a], Stephan Lany [a], John D. Perkins [a], Kevin R. Talley [b], and Andriy Zakutayev *[a]

[a]National Renewable Energy Laboratory, Golden, Colorado, USA, 80401
[b]Department of Metallurgical and Materials Engineering, Colorado School of Mines, Golden, Colorado, USA, 80401
*Corresponding authors

## Abstract

Nitrides feature many interesting properties, such as a wide range of bandgaps suitable for optoelectronic devices including light-emitting diodes (LEDs), and piezoelectric response used in microelectromechanical systems (MEMS). Nitrides are also significantly underexplored compared to oxides and other chemistries, with many being thermochemically metastable, sparking interest from a basic science point of view. This paper reports on experimental and computational exploration of the Mg-Sb-N material system, featuring both metastable materials and interesting semiconducting properties. Using sputter deposition, we discovered a new $Mg_2SbN_3$ nitride with a wurtzite-derived crystal structure and synthesized the antimonide-nitride $Mg_3SbN$ with an antiperovskite crystal structure for the first time in thin film form. Theoretical calculations indicate that $Mg_2SbN_3$ is metastable and has properties relevant to LEDs and MEMS, whereas $Mg_3SbN$ has a large dielectric constant ($28\varepsilon_0$) and low hole effective masses ($0.9m_0$), of interest for photovoltaic solar cell absorbers. The experimental solar-matched 1.3 eV optical absorption onset of the $Mg_3SbN$ antiperovskite agrees with the theoretical prediction (1.3 eV direct, 1.1 eV indirect), and with the measurements of room-temperature near-bandgap photoluminescence. These results make an important contribution towards understanding semiconductor properties and chemical trends in the Mg-Sb-N materials system, paving the way to future practical applications of these novel materials.

# 1. Introduction

Nitride materials are useful for a variety of semiconductor applications [1] [2], due to their intermediate ionic/covalent bonding character relative to other first-row compounds (i.e. oxides, fluorides) and other group-V compounds (i.e. phosphides, arsenides) [3]. However, there are over an order of magnitude fewer nitrides than oxides in crystallographic databases, such as ICSD [4] and ICDD [5]. Possible reasons are the strength of the nitrogen triple bond, making nitrides generally more difficult to synthesize than oxides, and the relative metal-nitrogen bond strength, making them more metastable than any other material family [6] [7]. As a result, there are a large number of nitride materials that have yet to be experimentally synthesized, many of which are predicted to have favorable optical, electronic, thermal, and mechanical properties [8] [9]. Several predicted ternary nitrides have recently been synthesized [10]. For some elements, such as Sb and Bi, even crystalline binary nitrides have not yet been reported.

Perovskite structures have been a known class of functional materials for a long time [11], and more recently have become an important area of solar cell research, due to their exceptional performance in photovoltaic devices [12] [13]. The discovery of a perovskite material that is stable upon exposure to heat, light, and bias, has a bandgap around the optimal point for absorption of sunlight, and is made of earth-abundant elements would be a breakthrough for photovoltaic research. Another property of interest in perovskites, for both functional materials and photovoltaic applications, is a large dielectric constant. This property is desirable for some applications of piezoelectrics and ferroelectrics [14] and can improve solar energy conversion efficiency through suppression of Shockley-Read-Hall (SRH) recombination by screening the electron-defect interaction [15].

A very limited number of nitride perovskites and Sb-containing nitrides have been reported. There are few existing experimental reports on oxygen-containing LaWN$_3$ [16] [17] and TaThN$_3$ [18] perovskites synthesized in bulk form, and a few recent theoretical papers on these and related materials [19] [20]. Several other nitride perovskites have been theoretically predicted [21] [22], and LaWN$_3$ was synthesized in the thin film form by our group [23]. There are also some bulk synthesis reports on antiperovskites, such as Mn$_3$*M*N (M=Cu, Zn, Ga, Sb) [24] [25] and *AE*$_3$SbN (AE = Mg, Ca, Sr, Ba) [26] [27], which are perovskites with the cation and anion sites reversed compared to the regular perovskite structure. This swapping of anions and cations is expected to lend itself to similar properties (e.g. large dielectric constants), with some electronic structure differences originating from elements swapping the roles of the cation vs. anion (e.g. Sb$^{3+}$ or Sb$^{5+}$ vs. Sb$^{3-}$). Beside the few known Sb-anion based antiperovskites, the first Sb-cation based nitride material Zn$_2$SbN$_3$ has been reported very recently by our group [28].

In this paper, the experimental discovery of a previously unreported antimony-based nitride material Mg$_2$SbN$_3$ with wurtzite-derived crystal structure is presented. According to first-principles theoretical calculations Mg$_2$SbN$_3$ is a metastable semiconductor, with synthesis enabled by activated nitrogen precursors. This paper also presents the first thin film synthesis and characterization of Mg$_3$SbN with antiperovskite structure (previously reported only in powder form), which has here enabled the reported electrical and optical properties. The results of the optical absorption, near-gap photoluminescence, and electrical conductivity measurements indicate that Mg$_3$SbN is a semiconductor with 1.3 eV bandgap, in reasonable agreement with 1.3 eV direct (1.1 eV indirect) bandgap predicted by theory. Theoretical calculations also indicate that Mg$_3$SbN has a large dielectric constant (28$\varepsilon_0$) and relatively low and well-matched effective masses of holes (0.9$m_0$) and electrons (0.7$m_0$), which are promising for semiconductor applications. Overall, this work makes an important step towards understanding of chemical relations and physical properties of semiconducting compounds in the Mg-Sb-N materials system.

## 2. Methods

*2.1 Synthesis*

Experimental synthesis of combinatorial Mg-Sb-N sample libraries was executed via co-sputtering of Mg and Sb metals in the presence of nitrogen and argon gasses. The Mg and Sb targets were both 2" diameter and were positioned at an angle 35º relative to the substrate normal, situated at diametrically opposite sides of the chamber, as depicted in Figure 1. This facilitates a composition gradient from Mg-rich to Sb-rich, with nitrogen throughout the sample. To access a wide range of Mg to Sb cation ratios across the sample libraries, the target power was varied between 40W and 60 W for the Mg target and kept constant at 20 W for the Sb target. For ease of further characterization, films were deposited on a variety of substrates depending on subsequent property measurements, including Corning Eagle XG glass and Delta Technologies $In_2O_3$:Sn (ITO) coated glass for ease of optical measurements on insulating and conductive substrates respectively, and GM associates fused silica plates or glassy carbon for compositional analysis.

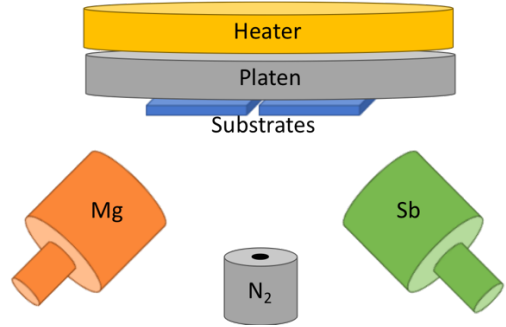

*Figure 1. Layout for the sputtering system. Mg and Sb targets mounted in RF sputtering guns provide the Mg and Sb components, respectively, and $N_2$ gas and Ar make up the remainder of the species entering the deposition system.*

The sputtering chamber had a base pressure of $2 \times 10^{-7}$ Torr. For the deposition, the sputtering pressure was set to 6 mTorr, with 6 sccm of Ar and 3 sccm of $N_2$ flowing into the system. To further explore the system, deposition pressure and relative Ar and $N_2$ flow rates were varied, though the optimal results were found at the parameters given above. A temperature gradient orthogonal to the applied compositional gradient was achieved by using a platen that made contact to one edge of the substrate while leaving the other end free. Thermal contact was made using silver paint, with the other end cantilevered out further from the platen heat sink between the heater and the substrate. This temperature gradient was applied to select substrates during growth to achieve a range of temperatures between 100 ºC and 510 ºC. When the temperature gradient platen was not used, the rows were all held at approximately the same temperature, due to a uniform back-side heating from the thermally conductive single-temperature platen.

Since nitrides tend to be more metastable than oxides, some of the films were predictably seen to react with oxygen in the air upon removal from the growth chamber. Films grown at conditions with a tendency to oxidize *ex-situ* were capped with a layer of aluminum nitride prior to removal from the system (grown *in-situ* immediately following the MgSbN deposition). The AlN capping layer prevents diffusion of oxygen into the crystalline MgSbN thin film, and the large band gap of AlN (and consequent transparency of the film) minimizes the shielding of optical properties. Thus, the determination of optical absorption coefficient of the thin film layers via optical spectroscopy was not significantly hindered by the capping layer. In addition, the AlN layer was well-aligned and well known, with the c-axis oriented along the substrate normal and the layer thickness set to around 100 nm. Growth conditions for AlN in this deposition system can be found elsewhere [29]. This protective capping layer increased the lifetime of the nitride sputtered samples from a few hours to a few months.

*2.2 Characterization*

Once the films were deposited, a grid of 4 x 11 points was used to characterize the samples, using the 11 columns as 11 different compositions across the spread of the wafer, and the 4 rows as 4 different temperatures when a temperature gradient platen was implemented. Various mapping characterization techniques were utilized across the combinatorial libraries to probe the structure-property relationships across the Mg-Sb-N composition space. Characterization results were managed, analysed, and displayed using the COMBIgor package for Igor Pro [30]. These techniques include structural and compositional measurements such as X-ray Diffraction (XRD) and Rutherford Backscattering Spectroscopy (RBS), respectively. Further study of the microstructural morphologies was done using Scanning Electron Microscopy (SEM) and both UV/Visible Spectroscopy and Photoluminescence (PL) were used to help determine the optical band gap of the material. To measure electrical conductivity, through-sample I-V measurements were performed.

For crystal structure determination of the materials grown across the Mg-Sb-N composition space, XRD was done using a Bruker D8 Discover XRD system with a Be window 2D detector. This lab x-ray source utilized primarily Cu K-$\alpha$ radiation, with a wavelength of 1.5406 Å. XRD was performed across the entirety of the combinatorial libraries for a finer examination of the crystal structure across the compositional gradient. For ease of analysis, the XRD patterns were then radially integrated to create a more conventional pseudo θ-2θ plot of intensity vs. scattering angle, as shown in the analysis section.

To determine the composition of the Mg-Sb-N, RBS was used on a few select samples within the libraries, since both magnesium and nitrogen are difficult or impossible to measure using more conventional methods such as X-ray Fluorescence (XRF). RBS data was analysed using the RUMP analysis software [31]. With the AlN capping layer deconvoluted from the Mg-Sb-N layer, the relative ratios of the elements in the film were determined. For calculations of composition in the Mg-Sb-N layer, the antimony content was taken to be a constant, set at 1, and the Mg and N compositions of the films were fitted relative to the Sb. The concentration of oxygen was determined through modelling of the layers with and without oxygen content and comparing the model to the measured data.

Further examination of the microstructural morphology of a few select MgSbN samples was done via both cross-sectional and plan-view SEM. SEM images were taken on a Nova 360 NanoSEM with a through-lens detector. The images were taken with a voltage of 2.00 kV and a current of 64 pA, with magnifications ranging from 5,000x to 100,000x. Both primarily $Mg_2SbN_3$ and primarily $Mg_3SbN$ films were examined.

For characterization of the optical properties, UV/Visible Spectroscopy was used in conjunction with Photoluminescence (PL), which together give a reasonable approximation of the optical bandgap of the material. To determine the optical bandgap, representative samples of the $Mg_3SbN$ antiperovskite phase were analysed via UV/visible spectroscopy. Transmission and reflection data were taken across a span of wavelengths ranging from near IR into the UV spectrum. Thickness of the films was measured via profilometry and used to calculate the absorption coefficient of the $Mg_3SbN$ film across the range of wavelengths. The PL data was taken for 5 seconds using a 632.8 nm HeNe laser at 5 mW power to pump the sample.

To analyse the electrical properties of the material and determine a resistivity of the $Mg_3SbN$ phase, through-sample IV measurements were used. The resistivity of the combinatorial libraries across a wide range of compositions was relatively high, lending these films to be more easily characterized via thru-film methods. Thus, the sample for electrical property analysis was grown on a conductive ITO on glass substrate, and a protective capping/contact stack of highly doped transparent ZnO was added ex-situ to masked off areas of the film. To ensure good contact through to the top surface of the sample, e-beam evaporation of 10 nm of Ti for adhesion and 100 nm Au as the top contact was added after the ZnO deposition. The backside contact to the ITO substrate was made by soldering an indium dot to a bare (masked off during the deposition) area of the ITO. The through-sample I-V measurement was taken through a film thickness of around 200 nm, with a contact area of 5 $mm^2$, and a voltage generally ranging from -1 V to +1 V, with testing up to ± 20 V to ensure that the contact resistance was not the dominant factor.

*2.3 Calculations*

The first-principles calculations in density functional theory (DFT) and in the GW approximation were performed with the VASP code [32] [33]. The compound enthalpy of formation was calculated using fitted elemental reference energies [34]. The GW calculations were performed as described in other work [35]. The optical absorption coefficient was calculated from the frequency dependent dielectric matrix in the independent particle approximation [36]. Density functional perturbation theory was used to calculate the electronic [36] and ionic [37] contributions to the static dielectric constant.

## 3. Results and Discussion

### *3.1 Composition and Structure*

X-Ray Diffraction (XRD) was used to determine the structure of the films, Rutherford Backscattering Spectroscopy (RBS) was used to determine the composition, and Scanning Electron Microscopy (SEM) was used to examine the microstructural morphology of the films. The results of the first two characterization techniques can be compared to each other to give a more thorough view of the compositional and structural makeup of each point across the sample library.

Table 1 presents the results of the RBS analysis of the endpoints of the combinatorial libraries between Mg-rich and Sb-rich extremes at two different Mg sputtering gun powers. An example of the RBS data obtained from one of the points across the Mg-Sb-N film composition space is shown in Figure S1 of the supplemental information. From the RBS data, the spread of Mg/(Mg+Sb) ratios ranges from 0.29 to 0.86. For 40 W Mg gun power, the stoichiometry of the Mg-rich side is pretty close to the predicted $Mg_2SbN_3$ stoichiometry of the wurtzite-derived structure. For 60 W Mg gun power, the Mg and N concentrations are both significantly increased, despite the expected Mg-rich and N-poor $Mg_3SbN$ stoichiometry. Thus, there may be an excess of nitrogen at this end of the composition range of the films. The measured nitrogen content of the films was quite high throughout the entire composition spread (N/(Sb+N) = 0.57 – 0.84), ranging from N:Sb = 1:1 to N:Sb = 5:1. This high N content may be related to RBS measurement error for light elements; alternatively it may be caused by some amorphous magnesium nitride in the films, in addition to the targeted ternary crystalline phases. Throughout the capped films, the maximum oxygen content was around 5 at. % of the anion content in the film, which is comparable to other sputtered nitrides that contain oxophylic elements like Mg [38] [39].

*Table 1 RBS results of the elemental ratios in Mg-Sb-N films. The approximate oxygen content in all films is 5 at.%*

| Mg Power [W] | Sb Power [W] | Sample endpoint [Mg, Sb] | Mg/(Mg+Sb) Ratio | N/(N+Sb) Ratio | Approx. Stoichiometry |
|---|---|---|---|---|---|
| 40 | 20 | Mg | 0.74 | 0.78 | $Mg_3SbN_3$ |
| 40 | 20 | Sb | 0.29 | 0.57 | $MgSb_2N_2$ |
| 60 | 20 | Mg | 0.86 | 0.84 | $Mg_6SbN_6$ |
| 60 | 20 | Sb | 0.55 | 0.66 | $MgSbN_2$ |

SEM images of the microstructural morphology of the samples were taken at representative locations on the sample libraries that had Mg:Sb compositions close to $Mg_2SbN_3$ and $Mg_3SbN$. Figure 2 shows top-down and cross-sectional SEM images of these materials. The top-down views in Figures 2a and 2c show a difference in grain size between the two different phases, with $Mg_3SbN$ having larger grains that $Mg_2SbN_3$. However, the overall morphology of these two materials seems to be pretty similar, despite the $Mg_3SbN$ sample being capped with AlN to prevent atmospheric oxidation of this magnesium-rich composition. Figure 2b shows slanted columnar structure of the grains in $Mg_2SbN_3$. This is likely due to the 35-degree angles used for the Mg and Sb sputtering guns, as shown in Figure 1. Figure 2d shows less of this slant for $Mg_3SbN$, but that may be due to a difference in angle of the cross-sectional SEM cut. The $Mg_3SbN$ sample in Figure 2c and 2d is capped with around 100 nm of AlN, which should not have a significant interaction volume in the SEM results, so the visible grain structure is largely due to the Mg$_3$SbN.

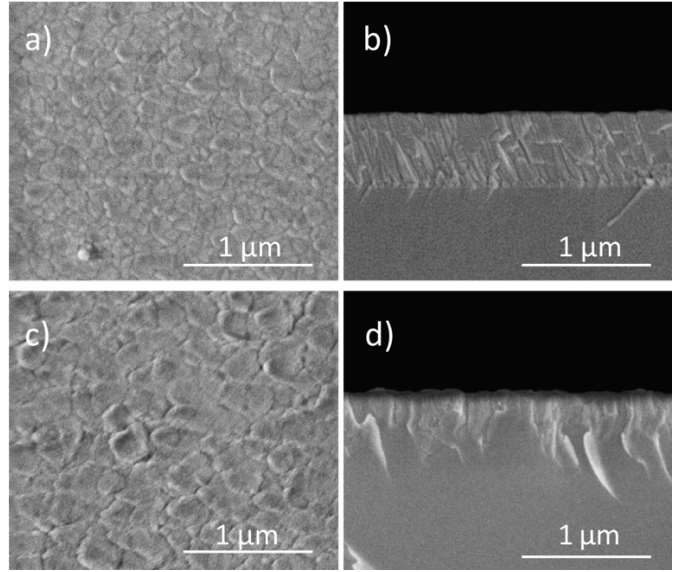

Figure 2. SEM images of sections of Mg$_2$SbN$_3$ and Mg$_3$SbN thin films a), b) are top-down and cross-sectional views of the wurtzite-derived Mg$_2$SbN$_3$ sample respectively; c), d) are top-down and cross-sectional views of the antiperovskite Mg$_3$SbN (capped with AlN), respectively

To measure the crystallinity of each of the grown films, and to determine the crystal structure by comparing to the reference patterns, XRD measurements were performed and the measured patterns are shown in Figure 3. The simulated diffraction peaks for various plane spacing between the atoms, determined from .cif files generated by theoretical calculations (available in the SI), are shown below the heat maps for both parts. Comparing these reference peaks with the peaks observed in the measured XRD heat map as a function of chemical composition (Figure 3a), at a higher Mg concentration these films are primarily in the Mg$_3$SbN antiperovskite structure, as evidenced by the peaks at approximately $2\theta = 35°$ (111) and $2\theta = 41°$ (200). As the Mg content decreases, the film transitions into the Mg$_2$SbN$_3$ orthorhombic (wurtzite-derived) phase as expected, indicated by the peaks at approximately $2\theta = 30°$ (310)/(020) and $2\theta = 35°$ (311)/(021)/(400). The peaks that show up at around $2\theta = 36°$ are from the (002) oriented AlN capping layer, used to lock in the nitrogen and prevent oxidation (particularly on the magnesium-rich side of the sample), as mentioned in the methods section.

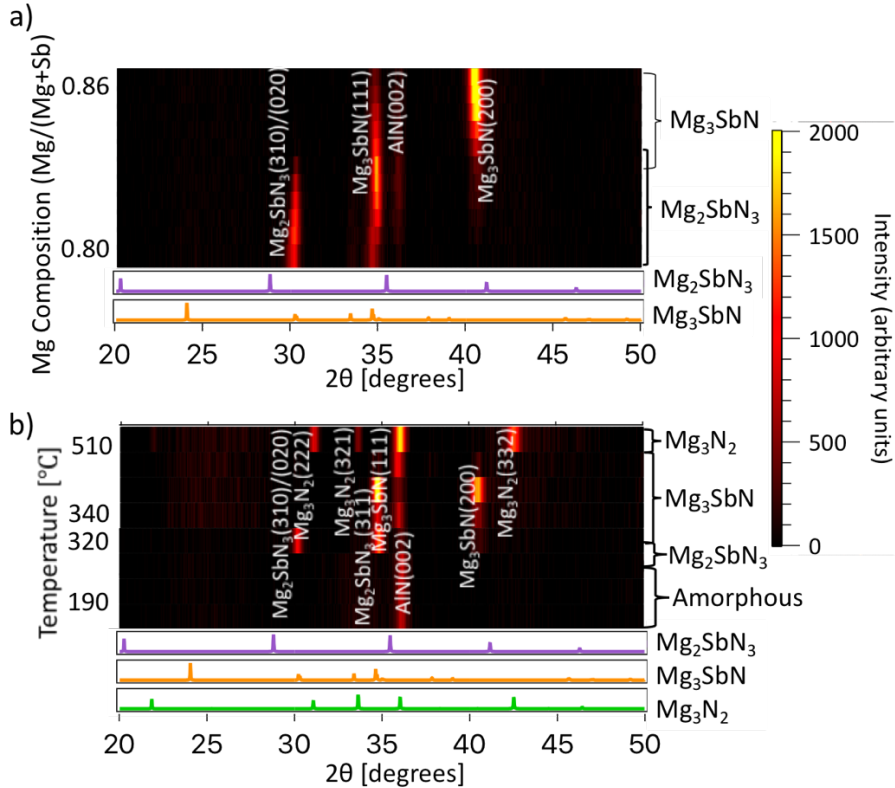

Figure 3. Color intensity XRD maps as a function of temperature and composition. Designations for each phase are directly to the left of the appropriate peak, and the phases are indicated along the same horizontal axis. a) Isothermal composition spread at T=320°C, showing $Mg_3SbN$ antiperovskite at higher Mg compositions, and $Mg_2SbN_3$ wurtzite-derived phase at lower Mg compositions. b) Isocompositional temperature spread at Mg/(Mg+Sb) = 0.82, showing the antiperovskite $Mg_3SbN$ phase at higher temperatures and wurtzite-like $Mg_2SbN_3$ phase at lower temperatures, with $Mg_3N_2$ and amorphous material at the highest and lowest temperatures respectively.

Figure 3b shows the XRD data across the temperature range of two sample libraries, at a particular concentration of Mg/(Mg+Sb) = 0.82. At the lowest temperatures, below approximately 300 ºC, only the AlN (002) peak from the capping layer is visible, indicating that the Mg-Sb-N film is amorphous. At approximately 320 ºC, the $Mg_2SbN_3$ orthorhombic structure is apparent with a peak appearing close to $2\theta = 30°$ (310)/(020). The peaks at $2\theta = 35°$ and $2\theta =$ 41° that appear in the range of temperatures between 320 ºC and 380 ºC are indicative of the Mg$_3$SbN antiperovskite structure, corresponding to the (111) and (200) peaks. At 510 ºC synthesis temperature, the Mg$_3$N$_2$ crystal structure dominates, potentially due to high Mg content caused by partial loss of Sb.

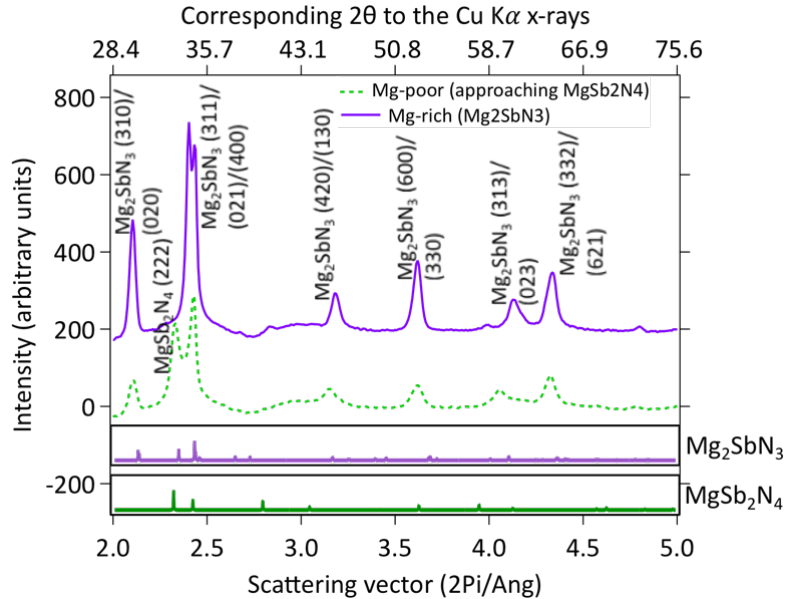

*Figure 4. Synchrotron-based XRD measurement results for the sample library straddling the Mg:Sb=1:1 composition. The Mg-rich Mg$_2$SbN$_3$ sample exhibits peaks of wurtzite-derived crystal structure, while in the Mg-poor sample there is also potential evidence of the MgSb$_2$N$_4$ phase with the peak around Q=2.3.*

To confirm the identification of the Mg$_2$SbN$_3$ with the orthorhombic crystal structure, synchrotron-based XRD measurements were performed at the Stanford Synchotron Radiation Lightsource (SSRL), beamline 1-5, of the SLAC National Laboratory [40]. Figure 4 shows selected XRD patterns taken for one of the Mg-Sb-N sample libraries that straddles the Mg:Sb = 1:1 composition ratio. At Mg-rich conditions, the Mg$_2$SbN$_3$ phase dominates, although preferential orientation of the films and cation disorder limit which peaks are visible. At the lowest measured Mg content, the MgSb$_2$N$_4$ phase may be present, although the Mg$_2$SbN$_3$ peaks are still dominant, and some of the peaks corresponding to these two structures overlap with each other. More synthesis and characterization experiments would be necessary to confirm or rule out the existence of the MgSb$_2$N$_4$ compound.

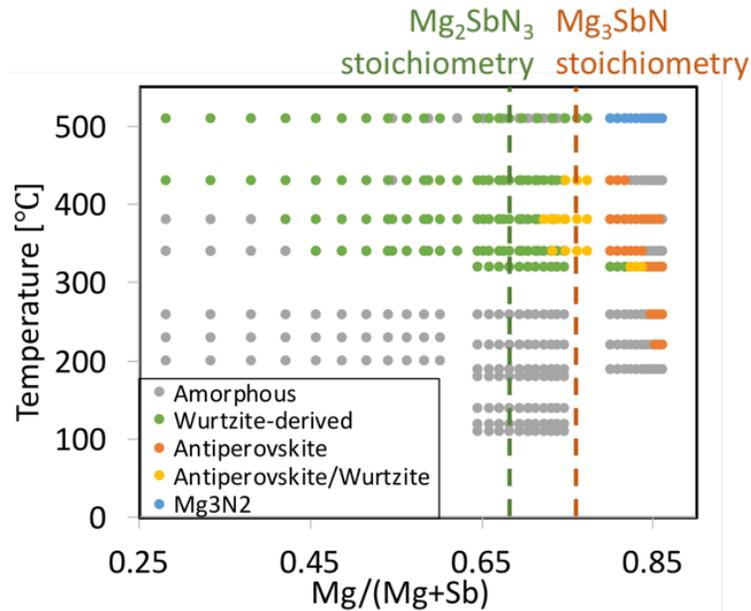

*Figure 5. Growth phase diagram for Mg-Sb-N materials systems, derived from the XRD measurements of the structure and RBS measurements of the composition, as a function of substrate temperature during synthesis. The locations for stoichiometric ratios between Mg and Sb for the Mg$_2$SbN$_3$ and the Mg$_3$SbN phases are shown.*

Combining the structure and composition data from XRD and RBS results in the growth phase diagram shown in Figure 5, which displays the resulting material structure as a function of synthesis temperature and Mg/(Mg+Sb) composition ratio for eight combinatorial sample libraries (352 sample points), with all other synthesis parameters fixed. The wurtzite-derived Mg$_2$SbN$_3$ structure crystallizes only above 300 ºC, and the Mg$_3$SbN antiperovskite structure disappears entirely above 500 ºC. Below 200 ºC growth temperature, all the films are amorphous, indicating that thermal energy is insufficient to crystallize these phases. In the intermediate temperature range (300 - 400 ºC), the transition from the antiperovskite phase to the wurtzite-derived phase occurs close to Mg/(Mg+Sb) = 0.82, which is higher than expected (0.67-0.75). This suggests that the resulting material may have some amorphous fraction of magnesium nitride.

*3.2 Computational Results*

Figure 6a shows the calculated Mg-Sb-N phase stability diagram as a function of the chemical potentials of all three elements. This representation is equivalent to the perhaps more common convex hull diagram, but it emphasizes the dependence on the synthesis conditions (chemical potentials) rather than the atomic ratios. The diagram includes both previously known phases ($Mg_3N_2$, $Mg_3Sb_2$, $Mg_3SbN$ [41]) and newly predicted phases ($SbN$, $Sb_3N_5$, $MgSb_2N_4$, $Mg_2SbN_3$), resulting from a computational ternary nitride stability search using prototype structures and unconstrained crystal structure prediction [8] [28] [10]. Numerous other stoichiometries were considered but did not turn out to be stable and therefore do not occur on the phase diagram. The predicted crystal structures of $MgSb_2N_4$ (spinel, space group Fd-3m), $Mg_2SbN_3$ (cation-ordered wurtzite-derived orthorhombic crystal, $Pmc2_1$), and $Mg_3SbN$ (anti-perovskite, Pm-3m) are shown in Fig. 6b. The orthorhombic structure $Mg_2SbN_3$ phase is derived from the hexagonal wurtzite structure by atomic ordering of the Mg and Sb on the cation sublattice. Thus, the orthorhombic cell (lattice parameters $a_o$, $b_o$, $c_o$) is a supercell of the wurtzite cell ($a_w$, $c_w$). It is notable that the ratio of the lattice parameters $a_o/b_o = 1.738$ is very close to $\sqrt{3} = 1.732$ for the ideal wurtzite supercell, making it difficult to distinguish between an ordered and a disordered cation arrangement using conventional x-ray diffraction [42] [43]. The predicted effective c/a ratio of 1.573 is considerably smaller than the ideal wurtzite value of 1.633. Overall, these computational results help to confirm the wurtzite-derived crystal structure of the new

Mg$_2$SbN$_3$ phase observed at intermediate chemical compositions in the Mg-Sb-N materials system (Figure 5).

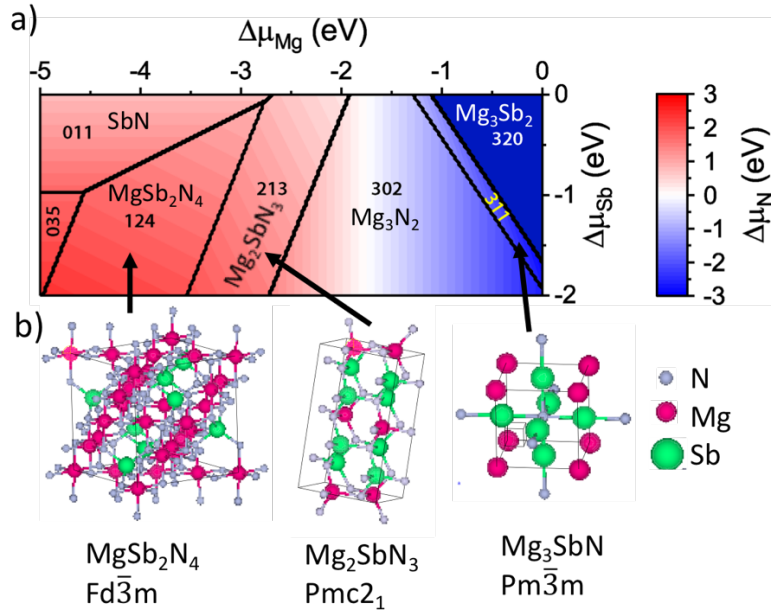

*Figure 6. (a)Calculated Mg-Sb-N chemical potential phase diagram. Each axis, including color, correspond to a compositional axis in a ternary phase diagram for these materials. (b) Ball and stick models for the three ternary phases of interest*

The systematic change in the Sb role and its valence state makes the Mg-Sb-N materials system interesting for studying chemical trends in properties of inorganic compounds. Furthermore, the computational phase diagram (Fig. 6a) can be used to guide the experimental synthesis. Under equilibrium growth, the N chemical potential can be easily converted by using the ideal gas law [44]. For non-equilibrium growth, our previous work has established a relationship between $\Delta\mu_N$ and process parameters [45]. We observe in Fig. 2a that the known phases (Mg$_3$Sb$_2$, Mg$_3$SbN) with the anionic state Sb$^{3-}$ exist under thermodynamically accessible negative chemical potentials $\Delta\mu_N$. In particular, the synthesis of Mg$_3$SbN requires only a very low nitrogen activity, -2.87 eV < $\Delta\mu_N$ < -0.71 eV (corresponding to $p$N$_2$ > 10$^{-16}$ atm at 800 °C [41]). For the previously unknown phases, stabilization of the cationic states of Sb, i.e., Sb$^{3+}$ in SbN and $Sb^{5+}$ in $Mg_2SbN_3$ and $MgSb_2N_4$, requires an activated non-equilibrium nitrogen source ($\Delta\mu_N > 0$). $Mg_2SbN_3$ requires only a slightly positive $\Delta\mu_N > +0.26$ eV, but the $MgSb_2N_4$ phase requires a fairly high N activation $\Delta\mu_N > +0.86$ eV, which is close to the upper limit for $\Delta\mu_N \approx 1$ eV estimated from previous studies [45] with a similar synthesis setup.

The calculated properties of $Mg_3SbN$ and $Mg_2SbN_3$ materials are presented in Table 2. The antiperovskite $Mg_3SbN$ phase is predicted to have a dielectric constant of $28\varepsilon_0$, higher even than that of halide perovskites ($22\varepsilon_0$) [46]. This high dielectric constant, together with an indirect bandgap of 1.1 eV and a direct bandgap of 1.3 eV, make $Mg_3SbN$ an interesting candidate for absorber applications in photovoltaic solar cells. The effective mass of holes and electrons in the $Mg_3SbN$ antiperovskite estimated from the density of states is $0.9m_0$ and $0.7m_0$, respectively.

Table 2 Calculated properties of the antiperovskite $Mg_3SbN$ and wurtzite-derived $Mg_2SbN_3$ phases in the Mg-Sb-N materials system.

| Crystal structure | Wurtzite-like | Antiperovskite |
|---|---|---|
| Chemical formula with valence states | $Mg_2SbN_3$ Sb(+5) cation | $Mg_3SbN$ Sb(-3) anion |
| Predicted bandgap and type | 3.1 eV, direct | 1.1 eV, indirect 1.3 eV, direct |
| Effective mass, electrons & holes | $m^*_e/m_0 = 0.4$, $m^*_h/m_0 \cong 2.0$ | $m^*_e/m_0 = 0.7$, $m^*_h/m_0 = 0.9$ |
| Dielectric Constant (static) | $8.7\varepsilon_0$ | $28\varepsilon_0$ |

The wurtzite-derived $Mg_2SbN_3$ phase with Sb cations [28] is predicted to have a much wider direct bandgap (3.1 eV) and a lower dielectric constant ($8.7\varepsilon_0$) than $Mg_3SbN$. The effective mass of electrons in the wurtzite-derived $Mg_2SbN_3$ is predicted to be $0.4m_0$, and the effective mass of holes is $2.0m_0$. The low electron effective mass and the wide direct band gap, which may be tuned down by degree of cation disorder [47] [48], makes this $Mg_2SbN_3$ material potentially interesting for light emitting diode (LED) applications in green or amber wavelength ranges where the efficiency of (In,Ga)N LEDs is lower [49] [50]. In addition, the non-centrosymmetric character of the wurtzite-derived structure of this material, similar to AlN and (Al,Sc)N alloys [51] [29], may be suitable for applications in piezoelectric microelectromechanical system (piezoMEMS) devices used in telecommunications.

### *3.3 Optical and Electrical Properties*

Optical and electrical properties of $Mg_3SbN$ antiperovskite were measured to validate the theoretical model for this material, and to indirectly support its predicted low effective masses and high dielectric constant. These experimental results can also give a preliminary estimate for the potential use of this material in photovoltaics or other optoelectronic applications.

To determine the optical bandgap, representative samples of $Mg_3SbN$ antiperovskite phase were measured by transmittance and reflectance spectroscopy, with the results shown in Figure S2 in the supplemental information. Combining the transmission (T) and reflection (R) data with the thickness (t) of the films determined via profilometry, the absorption coefficient of these materials was calculated via the relationship $\alpha = -\frac{1}{t}\ln\left(\frac{T}{1-R}\right)$ used for thin films. Figure 7 shows the measured absorption coefficient for $Mg_3SbN$ as a function of photon energy, along with the predicted theoretical absorption spectrum. Based on the $10^3$ cm$^{-1}$ cutoff point of the absorption coefficient, the optical bandgap of these $Mg_3SbN$ films is close to 1.3 eV, which matches the theoretical predictions for the direct band gap (Table 2).

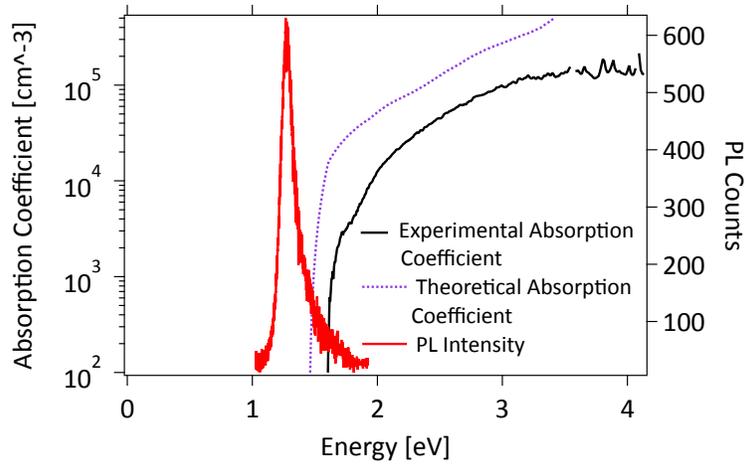

*Figure 7. Measured absorption coefficient for Mg$_3$SbN as a function of photon energy, alongside the predicted theoretical absorption plot. Right axis is the measured photoluminescence intensity in counts*

Room temperature photoluminescence of the Mg$_3$SbN antiperovskite phase is also shown in Figure 7. First of all, it is remarkable that this material sputtered on glass substrates at low temperatures shows room-temperature photoluminescence, which may be indicative of its defect tolerance due to the large dielectric constant. The maximum intensity of the photoluminescence peak close to 1.3 eV matches with the optical bandgap obtained through optical absorption spectroscopy, further confirming the theoretical bandgap estimate. This estimated bandgap of 1.3 eV from these experimental and theoretical results makes Mg$_3$SbN a good candidate for photovoltaic absorber applications, even with its slightly lower 1.1 eV indirect bandgap (Table 2).

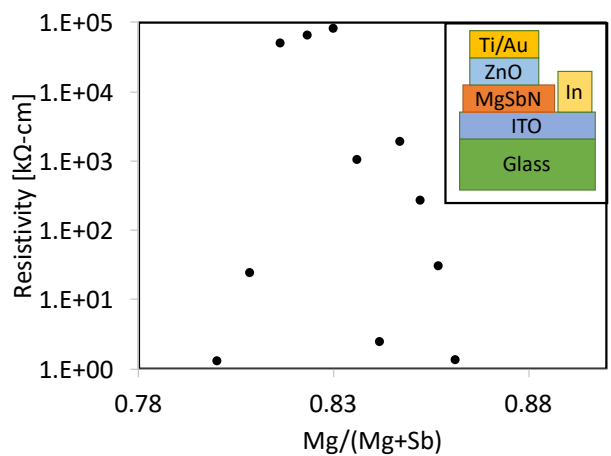

*Figure 8. Resistivity of the Mg-Sb-N thin films determined from IV measurements through the thickness of the sample, as a function of the Mg/(Mg+Sb) composition along the library. Inset is the stack used to measure the electrical properties.*

Electrical resistivity of the $Mg_3SbN$ films was determined from I-V measurement through the thickness of the sample. The resistivity of the sample with the strongest antiperovskite peaks at a composition of Mg/(Mg+Sb) = 0.83 was measured to be approximately 80 MΩ-cm. In the range of Mg/(Mg+Sb) composition of 0.80-0.86, the resistivity of the films varies significantly over several orders of magnitude. At the end points of the trend shown in Figure 8, the resistivity is close to 1-10 kΩ-cm, whereas in the middle of this composition spread the resistivity rises significantly to the order of 10-100 MΩ-cm. This significant increase in resistivity at stoichiometric composition supports the optical assessment that this material is a semiconductor that may be suitable for optoelectronic applications.

## 4. Summary and Conclusions

This research demonstrates several new and interesting findings in the Mg-Sb-N materials system. First, the $Mg_2SbN_3$ material with wurtzite-derived crystal structure was synthesized for the first time, and it was theoretically shown to be a metastable compound with a wide direct bandgap and low electron effective mass. Second, the $Mg_3SbN$ material with antipervoskite structure was deposited as a thin film for the first time, and its optoelectronic properties have been characterized. Through both optical absorption spectroscopy and photoluminescence measurements at room temperature, the optical bandgap of the antiperovskite $Mg_3SbN$ compound was determined to be around 1.3 eV, in good agreement with the theoretical predictions (1.3 eV direct, 1.1 eV indirect band gap). Theoretical calculations also indicate high dielectric constant ($28\varepsilon_0$) and low hole effective mass ($0.9m_0$) of this $Mg_3SbN$ antiperovskite material, both promising for the photovoltaic absorber applications. The overall growth phase diagram of the ternary Mg-Sb-N system has been mapped out as a function of synthesis temperatures and chemical composition. All these results make an important step towards scientific understanding of the Mg-Sb-N material structure and properties.

**Conflicts of interest**

There are no conflicts of interest to declare.

**Acknowledgements**

We thank Bobby To for SEM sample prep and images, Pat Dippo for taking the PL data, Karen Bowers for the doped ZnO deposition to aide in the electrical measurements, and Valerie Jacobson for her help with the SSRL data acquisition.

This work was supported by the U.S. Department of Energy under Contract No. DEAC36-08GO28308 with the Alliance for Sustainable Energy, LLC, the manager and operator of the National Renewable Energy Laboratory (NREL). The experimental work was funded by the Office of Energy Efficiency and Renewable Energy (EERE), under Solar Energy Technologies Office (SETO) Agreement Number 30302. The computational work was supported by the Office of Science, Basic Energy Sciences, as part of the Energy Frontier Research Center "Center for Next Generation of Materials Design: Incorporating Metastability". K.R.T. acknowledges funding support from the U.S. National Science Foundation (DMREF-1534503). This work used high-performance computing resources located at NREL and sponsored by DOE-EERE. The use of the Stanford Synchrotron Radiation Lightsource, SLAC National Accelerator Laboratory, is supported by DOE-SC-BES under contract No. DE-AC02-76SF00515. The views expressed in the article do not necessarily represent the views of the DOE or the U.S. Government.

## References

[1] H. Morkoc, Nitride Semiconductor Devices: Fundamentals and Applications, Wiley-VCH, 2013.

[2] J.-J. Huang, H.-C. Kuo and S.-C. Shen, Nitride Semiconductor Light-Emitting Diodes (LEDs): Materials, Technologies, and Applications, Woodhead Publishing, 2017.

[3] A. Zakutayev, "Design of nitride semiconductors for solar energy conversion," *Journal of Materials Chemistry A,* vol. 4, pp. 6742-6754, 2016.

[4] F. H. Allen and G. P. Shields, "Crystallographic Databases and Knowledge Bases in Materials Design," in *Implications of Molecular Materials Structure for New Technologies*, Springer, Dordrecht, 1999, pp. 291-302.

[5] "International Centre for Diffraction Data," Newtown Square, PA, USA, 2019.

[6] W. Sun, S. T. Dacek, S. P. Ong, G. Hautier, A. Jain, W. D. Richards, A. C. Gamst, K. A. Persson and G. Ceder, "The thermodynamic scale of inorganic crystalline metastability," *Science Advances,* vol. 2, 2016.

[7] W. Sun, A. Holder, B. Orvananos, E. Arca, A. Zakutayev, S. Lany and G. Ceder, "Thermodynamic Routes to Novel Metastable Nitrogen-Rich Nitrides," *Chemistry of Materials,* vol. 29, pp. 6936-6946, 2017.

[8] W. Sun, C. Bartel, E. Arca, S. Bauers, B. Matthews, B. Orvananos, B.-R. Chen, M. F. Toney, L. T. Schelhas, W. Tumas, J. Tate, A. Zakutayev, S. Lany, A. Holder and G. Ceder, "A Map of Inorganic Ternary Metal Nitrides," *Nature Materials,* 2019.

[9] Y. Hinuma, T. Hatakeyama, Y. Kumagai, L. A. Burton, H. Sato, Y. Muraba, S. Iimura, H. Hiramatsu, I. Tanaka, H. Hosono and F. Oba, "Discovery of earth-abundant nitride semiconductors by computational screening and high-pressure synthesis," *Nature Communications,* vol. 7, no. 11962, 2016.

[10] E. Arca, S. Lany, J. Perkins, C. Bartel, J. Magnum, W. Sun, A. Holder, G. Ceder, B. Gorman, G. Teeter, W. Tumas and A. Zakutayev, "Redox-Mediated Stabilization in Zinc Molybdenum Nitrides," *J. Am. Chem. Soc.,* vol. 140, no. 12, 2018.

[11] J. Bernard, W. R. Cook and H. Jaffe, Piezoelectric Ceramics, Elsevier, 1971.

[12] C. H. Ng, H. N. Lim, S. Hayase, Z. Zainal and N. M. Huang, "Photovoltaic performances of mono- and mixed-halide structures for perovskite solar cell: A review," *Renewable and Sustainable Energy Reviews,* vol. 90, pp. 248-274, 2018.

[13] Z. Li, T. R. Klein, D. H. Kim, M. Yang, J. J. Berry, M. F. A. M. van Hest and K. Zhu, "Scalable fabrication of perovskite solar cells," *Nature Reviews Materials,* vol. 3, 2018.

[14] R. J. D. Tilley, Perovskites: Structure-Property Relationships, Wiley, 2016.

[15] J. Brebels, J. V. Manca, L. Lutsen, D. Vanderzande and W. Maes, "High dielectric constant conjugated materials for organic photovoltaics," *Journal of Materials Chemistry A,* vol. 5, pp. 24037-24050, 2017.

[16] P. Bacher, P. Antoine, R. Marchand, P. L'Haridon, Y. Laurent and G. Roult, "Time-of-flight neutron diffraction study of the structure of the perovskite-type oxynitride $LaWO_{0.6}N_{2.4}$," *Journal of Solid State Chemistry,* vol. 77, no. 1, pp. 67-71, 1988.

[17] Y.-W. Fang, C. Fisher, A. Kuwabara, X.-W. Shen, T. Ogawa, H. Moriwake and R. Huang, "Lattice dynamics and ferroelectric properties of the nitride perovskite $LaWN_3$," *Physical Review B,* vol. 95, 2017.

[18] N. E. Brese and F. J. DiSalvo, "Synthesis of the First Thorium-Containing Nitride Perovskite, $TaThN_3$," *Journal of Solid State Chemistry,* vol. 120, pp. 378-380, 1995.

[19] J. M. Polfus and R. Haugsrud, "Protons in perovskite nitrides and oxide nitrides: A first principles study of $ThTaN_3$ and $SrTaO_2N$," *Solid State Communications,* vol. 152, no. 20, 2012.

[20] C. Paduani, "Electronic structure of the perovskite-type nitride $RuFe_3N$," *Journal of Magnetism and Magnetic Materials,* vol. 278, no. 1-2, 2004.

[21] J. A. Flores-Livas, R. Sarmiento-Perez, S. Botti, S. Goedecker and M. A. L. Marques, "Rare-earth magnetic nitride perovskites," *JPhys Materials,* vol. 2, 2019.

[22] R. Sarmiento-Perez, T. F. T. Cerqueira, S. Korbel, S. Botti and M. A. L. Marques, "Prediction of Stable Nitride Perovskites," *Chemistry of Materials,* vol. 27, pp. 5957-5963, 2015.

[23] K. Talley, J. Mangum, C. Perkins, R. Woods-Robinson, A. Mehta, . B. Gorman, G. L. Brennecka and A. Zakutayev, "Synthesis of lanthanum tungsten oxynitride perovskite thin films," *Advanced Electronic Materials,* 2019.

[24] M. Barberon, R. Madar, M. E. Fruchart, G. Lothioir and R. Fruchart, "Les deformations quadratiques T1 et T2 dans les carbures et nitrures perowskites du manganese," *Materials Research Bulletin,* vol. 5, no. 1, pp. 1-7, 1970.

[25] K. Takenaka, T. Shibayama, D. Kasugai and T. Shimizu, "Giant Field-Induced Distortion in $Mn_3SbN$ at Room Temperature," *Japanese Journal of Applied Physics,* vol. 51, 2012.

[26] K. Takenaka and H. Takagi, "Giant negative thermal expansion in Ge-doped anti-perovskite manganese nitrides," *Applied Physics Letters,* vol. 87, 2005.

[27] F. Gabler, M. Kirchner, W. Schnelle, U. Schwarz, M. Schmitt, H. Rosner and R. Niewa, "(Sr3N)E and (Ba3N)E (E = Sb, Bi): Synthesis, Crystal Structures, and Physical Properties," *S. Anorg. Allg. Chem.,* vol. 630, pp. 2292-2298, 2004.

[28] E. Arca, J. D. Perkins, S. Lany, A. Mis, B.-R. Chen, P. Dippo, J. L. Partridge, W. Sun, A. Holder, A. C. Tamboli, M. F. Toney, L. T. Schelhas, G. Ceder, W. Tumas, G. Teeter and A. Zakutayev, "Zn2SbN3: growth and characterization of a metastable photoactive semiconductor," *Materials Horizon,* 2019.

[29] K. R. Talley, S. L. Millican, J. Mangum, S. Siol, C. B. Musgrave, B. Gorman, A. M. Holder, A. Zakutayev and G. L. Brennecka, "Implications of heterostructural alloying for enhanced piezoelectric performance of (Al,Sc)N," *Physical Review Materials,* vol. 2, 2018.

[30] K. R. Talley, S. R. Bauers, C. L. Melamed, M. C. Papac, K. N. Heinselman, I. Khan, D. M. Roberts, V. Jacobson, A. Mis, G. L. Brennecka, J. D. Perkins and A. Zakutayev, "COMBIgor: data analysis package for combinatorial materials science," *ACS Combinatorial Science,* vol. Submitted, 2019.

[31] N. Barradas, K. Arstila, G. Battistig, M. Bianconi, N. Dytlewski, C. Jeynes, E. Kotai, G. Lulli, M. Mayer, E. Rauhala, E. Szilagyi and M. Thompson, "Summary of "IAEA intercomparison of IBA software"," *Nuclear Instruments and Methods in Physics Research, Section B: Beam Interactions with Materials and Atoms,* vol. 266, no. 8, pp. 1338-1342, 2008.

[32] G. Kresse and D. Joubert, "From ultrasoft pseudopotentials to the projector augmented-wave method," *Physical Review B,* vol. 59, pp. 1758-1775, 1999.

[33] M. Shishkin and G. Kresse, "Implementation and performance of the frequency-dependent GW method within the PAW framework," *Physical Review B,* vol. 74, 2006.

[34] V. Stevanovic, S. Lany, X. Zhang and A. Zunger, "Correcting density functional theory for accurate predictions of compound enthalpies of formation: Fitted elemental-phase reference energies," *Physical Review B,* vol. 85, 2012.

[35] S. Lany, "Band structure calculations for the 3d transition metal oxides in GW," *Physical Review B,* vol. 87, 2013.

[36] M. Gajdos, K. Hummer, G. Kresse, J. Furthmuller and F. Bechstedt, "Linear optical properties in the projector-augmented wave methodology," *Physical Review B,* vol. 73, 2006.

[37] X. Wu, D. Vanderbilt and D. R. Hamann, "Systematic treatment of displacements, strains, and electric fields in density-functional perturbation theory," *Physical Review B,* vol. 72, 2005.

[38] S. R. Bauers, D. M. Hamann, A. Patterson, J. D. Perkins, K. R. Talley and A. Zakutayev, "Composition, structure, and semiconducting properties of MgxZr2-xN2 thin films," *Japanese Journal of Applied Physics,* vol. 58, 2019.

[39] S. R. Bauers, A. Holder, W. Sun, C. M. Melamed, R. Woods-Robinson, J. Mangum, J. Perkins, W. Tumas, B. Gorman, A. Tamboli, G. Ceder, S. Lany and A. Zakutayev, "Ternary Nitride Semiconductors in the Rocksalt Crystal Structure," *arXiv,* 2019.

[40] J. M. Gregoire, D. G. Van Campen, C. E. Miller, R. J. R. Jones, S. K. Suram and A. Mehta, "High-throughput synchotron X-ray diffraction for combinatorial phase mapping," *Journal of Synchotron Radiation,* vol. 21, pp. 1262-1268, 2014.

[41] E. Chi, W. Kim, N. Hur and D. Jung, "New Mg-based antiperovskites PnNMg3 (Pn = As, Sb)," *Solid State Communications,* vol. 121, 2002.

[42] S. Lany, A. Fioretti, P. Zakwadzki, L. Schelhaus, E. Toberer, A. Zakutayev and A. Tamboli, "Monte Carlo simulations of disorder in ZnSnN2 and the effects on the electronic structure," *Physical Review Materials,* vol. 1, 2017.

[43] A. N. Fioretti, J. Pan, B. R. Ortiz, C. L. Melamed, P. C. Dippo, L. T. Schelhas, J. D. Perkins, D. Kuciauskas, S. Lany, A. Zakutayev, E. S. Toberer and A. C. Tamboli, "Exciton photoluminescence and benign defect complex formation in zinc tin nitride," *Materials Horizon,* vol. 5, pp. 823-830, 2018.

[44] J. Osorio-Guillen, S. Lany, S. V. Barabash and A. Zunger, "Magnetism without Magnetic Ions: Percolation, Exchange, and Formation Energies of Magnetism-Propoting Intrinsic Defects in CaO," *Physical Review Letters,* vol. 96, 2006.

[45] C. M. Caskey, R. M. Richards, D. S. Ginley and A. Zakutayev, "Thin film synthesis and properties of copper nitride, a metastable semiconductor," *Materials Horizons,* vol. 1, pp. 424-430, 2014.

[46] J. N. Wilson, J. M. Frost, S. K. Wallace and A. Walsh, "Dielectric and ferroic properties of metal halide perovskites," *APL Materials,* vol. 7, 2019.

[47] A. D. Martinez, A. N. Fioretti, E. S. Toberer and A. C. Tamboli, "Synthesis, structure, and optoelectronic properties of II-IV-V2 materials," *Journal of Materials Chemistry A,* vol. 5, pp. 11418-11435, 2017.

[48] N. Feldberg, J. D. Aldous, W. M. Linhart, L. J. Phillips, K. Durose, P. A. Stampe, R. J. Kennedy, D. O. Scanlon, G. Vardar, R. L. Field, T. Y. Jen, R. S. Goldman, T. D. Veal and S. M. Durbin, "Growth, disorder, and physical properties of ZnSnN2," *Applied Physics Letters,* vol. 103, 2013.

[49] F. A. Ponce and D. P. Bour, "Nitride-based semiconductors for blue and green light-emitting devices," *Nature,* vol. 386, pp. 351-359, 1997.

[50] T. Langer, A. Kruse, F. A. Ketzer, A. Schwiegel, L. Hoffman, H. Jonen, H. Bremers, U. Rossow and A. Hangleiter, "Origin of the "green gap": Increasing nonradiative recombination in indium-rich GaInN/GaN quantum well structures," *Physica Status Solidi C,* vol. 8, no. 7-8, pp. 2170-2172, 2011.

[51] M. Akiyama, T. Kamohara, K. Kano, A. Teshigahara, Y. Takeuchi and N. Kawahara, "Enhancement of Piezoelectric Response in Scandium Aluminum Nitride Alloy Thin Films Prepared by Dual Reactive Cosputtering," *Advanced Materials,* vol. 21, pp. 593-596, 2009.

# Supplemental Information for Thin film synthesis of semiconductors in the Mg-Sb-N materials system

Karen N. Heinselman *[a], Stephan Lany [a], John D. Perkins [a], Kevin R. Talley [b], and Andriy Zakutayev *[a]

[a]National Renewable Energy Laboratory, Golden, Colorado, USA, 80401
[b]Department of Metallurgical and Materials Engineering, Colorado School of Mines, Golden, Colorado, USA, 80401
*Corresponding authors

**Supplemental Information**

RBS data was taken on several points across the Mg to Sb spread in a few of the samples. An example of the RBS data taken and the corresponding fit are shown in Figure S1.

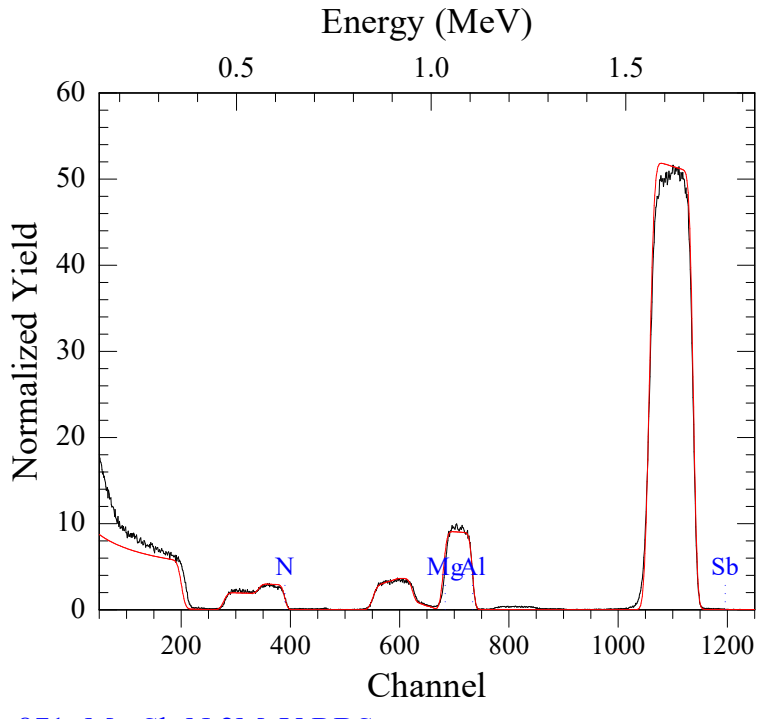

*Figure S1. Example RBS data and modelling to determine Mg:Sb:N:O ratios within the thin films*

Transmission and reflection data, along with the thickness of the films, was used to determine the absorption coefficient of the Mg-Sb-N films. An example of the transmission and reflection data taken is shown in Figure S2.

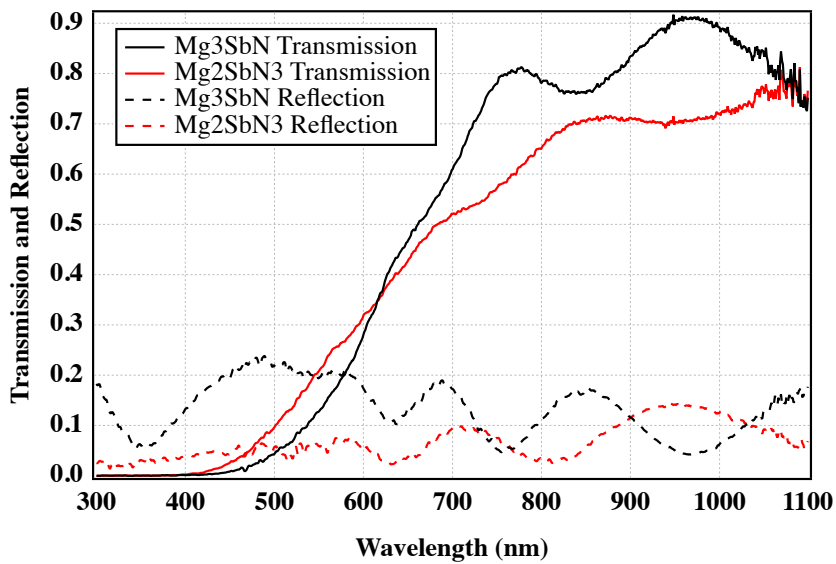

*Figure S2. Transmission and Reflection data on regions of primarily 311 and primarily 213 phase Mg-Sb-N.*

.cif file text

MgSb$_2$N$_4$

# CIF file
# This file was generated by FINDSYM
# Harold T. Stokes, Branton J. Campbell, Dorian M. Hatch
# Brigham Young University, Provo, Utah, USA

data_findsym-output
_audit_creation_method FINDSYM

_symmetry_space_group_name_H-M "F 41/d -3 2/m (origin choice 2)"
_symmetry_Int_Tables_number 227

_cell_length_a   8.99876
_cell_length_b   8.99876
_cell_length_c   8.99876
_cell_angle_alpha 90.00000
_cell_angle_beta  90.00000
_cell_angle_gamma 90.00000

loop_
_space_group_symop_id
_space_group_symop_operation_xyz
1 x,y,z
2 x,-y+1/4,-z+1/4
3 -x+1/4,y,-z+1/4
4 -x+1/4,-y+1/4,z
5 y,z,x
6 y,-z+1/4,-x+1/4
7 -y+1/4,z,-x+1/4

8 -y+1/4,-z+1/4,x
9 z,x,y
10 z,-x+1/4,-y+1/4
11 -z+1/4,x,-y+1/4
12 -z+1/4,-x+1/4,y
13 -y,-x,-z
14 -y,x+1/4,z+1/4
15 y+1/4,-x,z+1/4
16 y+1/4,x+1/4,-z
17 -x,-z,-y
18 -x,z+1/4,y+1/4
19 x+1/4,-z,y+1/4
20 x+1/4,z+1/4,-y
21 -z,-y,-x
22 -z,y+1/4,x+1/4
23 z+1/4,-y,x+1/4
24 z+1/4,y+1/4,-x
25 -x,-y,-z
26 -x,y+1/4,z+1/4
27 x+1/4,-y,z+1/4
28 x+1/4,y+1/4,-z
29 -y,-z,-x
30 -y,z+1/4,x+1/4
31 y+1/4,-z,x+1/4
32 y+1/4,z+1/4,-x
33 -z,-x,-y
34 -z,x+1/4,y+1/4
35 z+1/4,-x,y+1/4
36 z+1/4,x+1/4,-y
37 y,x,z
38 y,-x+1/4,-z+1/4
39 -y+1/4,x,-z+1/4
40 -y+1/4,-x+1/4,z
41 x,z,y
42 x,-z+1/4,-y+1/4
43 -x+1/4,z,-y+1/4
44 -x+1/4,-z+1/4,y
45 z,y,x
46 z,-y+1/4,-x+1/4
47 -z+1/4,y,-x+1/4
48 -z+1/4,-y+1/4,x
49 x,y+1/2,z+1/2
50 x,-y+3/4,-z+3/4
51 -x+1/4,y+1/2,-z+3/4
52 -x+1/4,-y+3/4,z+1/2
53 y,z+1/2,x+1/2
54 y,-z+3/4,-x+3/4
55 -y+1/4,z+1/2,-x+3/4
56 -y+1/4,-z+3/4,x+1/2
57 z,x+1/2,y+1/2
58 z,-x+3/4,-y+3/4
59 -z+1/4,x+1/2,-y+3/4
60 -z+1/4,-x+3/4,y+1/2
61 -y,-x+1/2,-z+1/2
62 -y,x+3/4,z+3/4
63 y+1/4,-x+1/2,z+3/4
64 y+1/4,x+3/4,-z+1/2
65 -x,-z+1/2,-y+1/2
66 -x,z+3/4,y+3/4
67 x+1/4,-z+1/2,y+3/4
68 x+1/4,z+3/4,-y+1/2
69 -z,-y+1/2,-x+1/2

70 -z,y+3/4,x+3/4
71 z+1/4,-y+1/2,x+3/4
72 z+1/4,y+3/4,-x+1/2
73 -x,-y+1/2,-z+1/2
74 -x,y+3/4,z+3/4
75 x+1/4,-y+1/2,z+3/4
76 x+1/4,y+3/4,-z+1/2
77 -y,-z+1/2,-x+1/2
78 -y,z+3/4,x+3/4
79 y+1/4,-z+1/2,x+3/4
80 y+1/4,z+3/4,-x+1/2
81 -z,-x+1/2,-y+1/2
82 -z,x+3/4,y+3/4
83 z+1/4,-x+1/2,y+3/4
84 z+1/4,x+3/4,-y+1/2
85 y,x+1/2,z+1/2
86 y,-x+3/4,-z+3/4
87 -y+1/4,x+1/2,-z+3/4
88 -y+1/4,-x+3/4,z+1/2
89 x,z+1/2,y+1/2
90 x,-z+3/4,-y+3/4
91 -x+1/4,z+1/2,-y+3/4
92 -x+1/4,-z+3/4,y+1/2
93 z,y+1/2,x+1/2
94 z,-y+3/4,-x+3/4
95 -z+1/4,y+1/2,-x+3/4
96 -z+1/4,-y+3/4,x+1/2
97 x+1/2,y,z+1/2
98 x+1/2,-y+1/4,-z+3/4
99 -x+3/4,y,-z+3/4
100 -x+3/4,-y+1/4,z+1/2
101 y+1/2,z,x+1/2
102 y+1/2,-z+1/4,-x+3/4
103 -y+3/4,z,-x+3/4
104 -y+3/4,-z+1/4,x+1/2
105 z+1/2,x,y+1/2
106 z+1/2,-x+1/4,-y+3/4
107 -z+3/4,x,-y+3/4
108 -z+3/4,-x+1/4,y+1/2
109 -y+1/2,-x,-z+1/2
110 -y+1/2,x+1/4,z+3/4
111 y+3/4,-x,z+3/4
112 y+3/4,x+1/4,-z+1/2
113 -x+1/2,-z,-y+1/2
114 -x+1/2,z+1/4,y+3/4
115 x+3/4,-z,y+3/4
116 x+3/4,z+1/4,-y+1/2
117 -z+1/2,-y,-x+1/2
118 -z+1/2,y+1/4,x+3/4
119 z+3/4,-y,x+3/4
120 z+3/4,y+1/4,-x+1/2
121 -x+1/2,-y,-z+1/2
122 -x+1/2,y+1/4,z+3/4
123 x+3/4,-y,z+3/4
124 x+3/4,y+1/4,-z+1/2
125 -y+1/2,-z,-x+1/2
126 -y+1/2,z+1/4,x+3/4
127 y+3/4,-z,x+3/4
128 y+3/4,z+1/4,-x+1/2
129 -z+1/2,-x,-y+1/2
130 -z+1/2,x+1/4,y+3/4
131 z+3/4,-x,y+3/4

132 z+3/4,x+1/4,-y+1/2
133 y+1/2,x,z+1/2
134 y+1/2,-x+1/4,-z+3/4
135 -y+3/4,x,-z+3/4
136 -y+3/4,-x+1/4,z+1/2
137 x+1/2,z,y+1/2
138 x+1/2,-z+1/4,-y+3/4
139 -x+3/4,z,-y+3/4
140 -x+3/4,-z+1/4,y+1/2
141 z+1/2,y,x+1/2
142 z+1/2,-y+1/4,-x+3/4
143 -z+3/4,y,-x+3/4
144 -z+3/4,-y+1/4,x+1/2
145 x+1/2,y+1/2,z
146 x+1/2,-y+3/4,-z+1/4
147 -x+3/4,y+1/2,-z+1/4
148 -x+3/4,-y+3/4,z
149 y+1/2,z+1/2,x
150 y+1/2,-z+3/4,-x+1/4
151 -y+3/4,z+1/2,-x+1/4
152 -y+3/4,-z+3/4,x
153 z+1/2,x+1/2,y
154 z+1/2,-x+3/4,-y+1/4
155 -z+3/4,x+1/2,-y+1/4
156 -z+3/4,-x+3/4,y
157 -y+1/2,-x+1/2,-z
158 -y+1/2,x+3/4,z+1/4
159 y+3/4,-x+1/2,z+1/4
160 y+3/4,x+3/4,-z
161 -x+1/2,-z+1/2,-y
162 -x+1/2,z+3/4,y+1/4
163 x+3/4,-z+1/2,y+1/4
164 x+3/4,z+3/4,-y
165 -z+1/2,-y+1/2,-x
166 -z+1/2,y+3/4,x+1/4
167 z+3/4,-y+1/2,x+1/4
168 z+3/4,y+3/4,-x
169 -x+1/2,-y+1/2,-z
170 -x+1/2,y+3/4,z+1/4
171 x+3/4,-y+1/2,z+1/4
172 x+3/4,y+3/4,-z
173 -y+1/2,-z+1/2,-x
174 -y+1/2,z+3/4,x+1/4
175 y+3/4,-z+1/2,x+1/4
176 y+3/4,z+3/4,-x
177 -z+1/2,-x+1/2,-y
178 -z+1/2,x+3/4,y+1/4
179 z+3/4,-x+1/2,y+1/4
180 z+3/4,x+3/4,-y
181 y+1/2,x+1/2,z
182 y+1/2,-x+3/4,-z+1/4
183 -y+3/4,x+1/2,-z+1/4
184 -y+3/4,-x+3/4,z
185 x+1/2,z+1/2,y
186 x+1/2,-z+3/4,-y+1/4
187 -x+3/4,z+1/2,-y+1/4
188 -x+3/4,-z+3/4,y
189 z+1/2,y+1/2,x
190 z+1/2,-y+3/4,-x+1/4
191 -z+3/4,y+1/2,-x+1/4
192 -z+3/4,-y+3/4,x

```
loop_
_atom_site_label
_atom_site_type_symbol
_atom_site_symmetry_multiplicity
_atom_site_Wyckoff_label
_atom_site_fract_x
_atom_site_fract_y
_atom_site_fract_z
_atom_site_occupancy
Mg1 Mg   8 b 0.37500 0.37500 0.37500 1.00000
Sb1 Sb  16 c 0.00000 0.00000 0.00000 1.00000
N1  N   32 e 0.76076 0.76076 0.76076 1.00000
```

# $Mg_2SbN_3$

```
# CIF file
# This file was generated by FINDSYM
# Harold T. Stokes, Branton J. Campbell, Dorian M. Hatch
# Brigham Young University, Provo, Utah, USA

data_findsym-output
_audit_creation_method FINDSYM

_symmetry_space_group_name_H-M "C m c 21"
_symmetry_Int_Tables_number 36

_cell_length_a    10.24663
_cell_length_b    5.89463
_cell_length_c    5.36251
_cell_angle_alpha 90.00000
_cell_angle_beta  90.00000
_cell_angle_gamma 90.00000

loop_
_space_group_symop_id
_space_group_symop_operation_xyz
1 x,y,z
2 -x,-y,z+1/2
3 -x,y,z
4 x,-y,z+1/2
5 x+1/2,y+1/2,z
6 -x+1/2,-y+1/2,z+1/2
7 -x+1/2,y+1/2,z
8 x+1/2,-y+1/2,z+1/2

loop_
_atom_site_label
_atom_site_type_symbol
_atom_site_symmetry_multiplicity
_atom_site_Wyckoff_label
_atom_site_fract_x
_atom_site_fract_y
_atom_site_fract_z
_atom_site_occupancy
Mg1 Mg   8 b 0.82982 0.16208 0.19908 1.00000
Sb1 Sb   4 a 0.00000 0.32442 0.69236 1.00000
N1  N    8 b 0.83872 0.17397 0.58736 1.00000
N2  N    4 a 0.00000 0.34539 0.07447 1.00000
```

# Mg$_3$SbN

```
# CIF file
# This file was generated by FINDSYM
# Harold T. Stokes, Branton J. Campbell, Dorian M. Hatch
# Brigham Young University, Provo, Utah, USA

data_findsym-output
_audit_creation_method FINDSYM

_symmetry_space_group_name_H-M "P 4/m -3 2/m"
_symmetry_Int_Tables_number 221

_cell_length_a    4.38213
_cell_length_b    4.38213
_cell_length_c    4.38213
_cell_angle_alpha 90.00000
_cell_angle_beta  90.00000
_cell_angle_gamma 90.00000

loop_
_space_group_symop_id
_space_group_symop_operation_xyz
1 x,y,z
2 x,-y,-z
3 -x,y,-z
4 -x,-y,z
5 y,z,x
6 y,-z,-x
7 -y,z,-x
8 -y,-z,x
9 z,x,y
10 z,-x,-y
11 -z,x,-y
12 -z,-x,y
13 -y,-x,-z
14 -y,x,z
15 y,-x,z
16 y,x,-z
17 -x,-z,-y
18 -x,z,y
19 x,-z,y
20 x,z,-y
21 -z,-y,-x
22 -z,y,x
23 z,-y,x
24 z,y,-x
25 -x,-y,-z
26 -x,y,z
27 x,-y,z
28 x,y,-z
29 -y,-z,-x
30 -y,z,x
31 y,-z,x
32 y,z,-x
33 -z,-x,-y
34 -z,x,y
35 z,-x,y
36 z,x,-y
37 y,x,z
38 y,-x,-z
```

```
39 -y,x,-z
40 -y,-x,z
41 x,z,y
42 x,-z,-y
43 -x,z,y
44 -x,-z,y
45 z,y,x
46 z,-y,-x
47 -z,y,-x
48 -z,-y,x

loop_
_atom_site_label
_atom_site_type_symbol
_atom_site_symmetry_multiplicity
_atom_site_Wyckoff_label
_atom_site_fract_x
_atom_site_fract_y
_atom_site_fract_z
_atom_site_occupancy
Sb1  Sb   1 a 0.00000 0.00000 0.00000 1.00000
Mg1  Mg   3 c 0.00000 0.50000 0.50000 1.00000
N1   N    1 b 0.50000 0.50000 0.50000 1.00000
```